\documentclass[a4paper,fleqn,usenatbib]{mnras}

\usepackage[T1]{fontenc}
\usepackage{ae,aecompl}

\usepackage{graphicx}	
\usepackage{epstopdf}	
\usepackage{amsmath}	

\title[MORSE code for GPUs]{Microlensing Observations Rapid Search for Exoplanets:  MORSE code for GPUs}

\author[Alistair McDougall and Michael D. Albrow]
{Alistair McDougall\thanks{E-mail:Alistair.McDougall@gmail.com}  and Michael D. Albrow\thanks{E-mail:Michael.Albrow@canterbury.ac.nz}
\ \\
Department of Physics and Astronomy, University of Canterbury, Private Bag 4800, Christchurch, \\ New Zealand 
}

\date{Accepted XXX. Received YYY; in original form ZZZ}

\pubyear{2015}

\begin{document}
\label{firstpage}
\pagerange{\pageref{firstpage}--\pageref{lastpage}}
\maketitle

\begin{abstract}
The rapid analysis of ongoing gravitational microlensing events has been integral to the successful
detection and characterisation of cool planets orbiting low mass stars in the Galaxy.
In this paper we present an implementation of search and fit techniques on Graphical Processing Unit 
hardware. The method allows for the rapid identification of candidate planetary microlensing events
and their subsequent followup for detailed characterisation.
\end{abstract}


\section{Introduction}

The OGLE\footnote{http://ogle.astrouw.edu.pl/ogle4/ews/ews.html}, MOA\footnote{https://it019909.massey.ac.nz/moa/alert2013/alert.html} 
and KMTNet\footnote{http://astroph.chungbuk.ac.kr/$\sim$~kmtnet}
wide-field surveys currently detect and alert in excess of 2000 microlensing events per year in the direction of  the Galactic Bulge. These events have an average timescale of $\sim$~22 days, but individual events can be as short as a few days or as long as a few hundred. A small subset of microlensing events produce anomalous short-duration signals lasting from a few hours to a few days due to the presence of one or more planets associated with the lens-star (which is typically an M dwarf). Almost all the planets that have been detected and characterised via microlensing have relied on high-cadence follow-up observations from various groups that include $\mu$FUN \citep{Gould2006}, RoboNet \citep{Tsapras2009}, PLANET \citep{Albrow1996,Albrow1998} and MiNDSTEp \citet{Dominik2010}. 

The follow-up groups typically operate with small-field-of-view cameras, capable of observing only one event at a time, so 
event selection can have a significant effect on the yield of microlensing planets.
The follow-up groups typically focus on two classes of events. First, \citet{Griest1998} have shown that, given the presence of a planet 
orbiting a lens star, a trajectory having a low impact parameter (i.e. a high magnification) is significantly more likely to intercept a
central caustic and therefore produce an anomalous signal. Thus events predicted to reach very high magnification
are  chosen for intensive observation over their peak magnification times.
Second, events that are observed to be undergoing an anomaly are priority targets for follow-up.
Unfortunately this second category of anomalous events is dominated by binary stars, rather than star-planet
microlensing. 

In order to select events from this category for intensive follow-up, rapid preliminary analysis of in-progress events 
is necessary. In this paper, we describe a new GPU-based code designed for rapid analysis of ongoing microlensing 
events. 

The layout of the paper is as follows. In Section~\ref{Section:GPU_architecture} we describe briefly the GPU computing architecture
on which the code is based. In Sections~\ref{Section:Microlensing_model}-\ref{Section:Refining_solutions} we present the algorithms and implementation
details. Section~\ref{Section:Examples} shows two examples of event analysis, and a summary is given in Section~\ref{Section:Summary}.

\section{GPU architecture}
\label{Section:GPU_architecture}

A Graphics Processing Unit (GPU) is a common component found in most desktop computers, but only recently has the potential of these devices for general computation been realized, in a range of disciplines from sciences to finance \citep{GPUfinance, GPUcosmology, GPUchemistry, GPUbiology}. By utilizing the large scale parallelization ability of GPUs, increased performance can be achieved for some computational calculations if correctly incorporated.
Most importantly, GPU code is generally inherently scalable, and increases in the number of GPU cores per device continues to outpace developments in
CPU technology (see, for instance, \citet{Vernardos2014}).

For this code, we have adopted the Compute Unified Device Architecture (CUDA) extensions to the C language, developed by NVIDIA Corporation. A CUDA-enabled device can be thought of 
as a grid of processing units (CUDA cores) that can execute sets of instructions (threads) in parallel. Blocks of up to 1024 threads (on compute capability 2.0) can be coded to access 
an amount of shared memory, with different threads (within a block) able to access individual elements of an array. Blocks can execute on the GPU either sequentially or in parallel, depending
on their size and the number of computing cores in the device. At the device level, groups of 32 threads within a block (warps) have hardware enforced parallel execution.
Blocks of threads and grids of blocks can have either 1-, 2- or 3-dimensional configurations (in CUDA compute capability 2.0 devices).

Additional to the shared memory (with fast read-write performance), threads have access to a slower read-write access global memory, and 
a hardware-optimised read-only texture memory. Both of these may be accessed by all threads, in all blocks of a single GPU function call.
Texture memory is designed to store 2- and 3-dimensional arrays, and the computing architecture is designed for very rapid
access to neighbouring pixels. This allows, for instance, very fast bilinear interpolation of an image stored in texture memory.

\section{The binary microlensing model}
\label{Section:Microlensing_model}

The characteristic angular scale for microlensing is the Einstein radius,
 \begin{equation}
\theta_E = \sqrt{\frac{4 G M}{c^2} \frac{D_S - D_L}{D_S D_L}},
\end{equation}
where $M$ is the total lens mass and $D_S$ and $D_L$ are the
distances to the source and the lens respectively.

The lensing equation for a point source and binary point-mass lens can be written
\begin{equation}
\zeta = z -  \frac{\varepsilon_1} {\bar{z}-\bar{z}_{1}} - \frac{\varepsilon_2} {\bar{z}-\bar{z}_{2}},	
\label{binary_eqn_6}
\end{equation}
where $\varepsilon_1$ and $\varepsilon_2$ are the masses of the individual lens components (as a proportion of the total lens mass),
$z_1$ and $z_2$ are their complex angular positions in the plane of the sky in units of the Einstein radius \citep{Witt1995}. $\zeta$ is the source position and 
$z$ is the image position.
Inversion of this equation gives a fifth-order complex polynomial for the image positions, with either 3 or 5 solutions.

Since surface brightness is preserved by gravitational lensing, the magnification of the source in image $j$,
\begin{equation}
A_j = \frac{1}{det J},
\end{equation}
where 
\begin{equation}
J = \frac{\partial{\zeta}}{\partial{z}}
\end{equation}
is the Jacobian of the coordinate transformation from the source to the image plane.
The total magnification is the sum of the individual image magnifications.

The point-source model for a microlensing event is rarely adequate for the majority of 
recognised binary microlensing events. In these cases, the source tends to pass close-to, or over, 
a lens caustic, which serves to resolve the source. To model such events, we must use a methodology that
accounts for the finite source size. The two most common algorithms for this are contour integration via Stokes'
Theorem \citep{Gould1997, Bozza2010} and various refinements of 
ray shooting maps \citep{Kayser1986, Dong2006, Bennett1996}. In Sections~\ref{Section:Coarse_grid_search}-\ref{Section:Refining_solutions}
below we describe our strategy, that is based on the magnification map approach.

\section{Coarse grid search}
\label{Section:Coarse_grid_search}

\subsection{Magnification maps}

Adopting Einstein-radius-normalized coordinates, set in the lens plane with the 
origin as the centre of mass
of the lens and the lens masses lying on the $z_1$ axis, 
a binary microlensing event may be described by seven basic parameters:
$d$, the angular separation of the two lens components; $q$, the mass ratio
of the lens components; $u_0$ the distance of closest approach of the source to lens' barycentre; 
$\phi$, the trajectory angle of the source measured from the positive axis that passes through both lens components; $t_0$ the time of
closest approach; $t_E$, the time for the source to move an angular distance of one Einstein radius;
and $\rho$, the angular source radius.

The majority of ground-based microlensing photometry is derived using the difference-imaging method.
The flux of a microlensing event,
\begin{equation}
F_i = \Delta F_i + F_{\rm Ref} = F_S A(t_i) + F_B, 
\end{equation}
where $F_{\rm Ref}$ is the flux of the star on the reference image, $F_S$ is the unlensed source star flux, $F_B$ is
the flux due to blended objects and $\Delta F_i$ is the measured difference-image flux. 
The difference-flux model is therefore a linear function of the magnification,
\begin{equation}
\Delta F_i = c_0 + c_1 A(t_i),
\end{equation}
where $A(t)$ is itself a non-linear function of the fundamental parameters $(t_0, t_E, u_0, \phi, \rho, d, q)$. 
The first part of our modelling method is a search for viable regions of solution space in $(t_0, t_E, u_0, \phi, \rho, d, q)$.

At the highest level, we set up a $29\times21$ fixed grid in $(\log d, \log q)$. 
For each $(\log d,\log q)$ pair, we generate a point-source magnification
map, by solving the fifth-order polynomial inversion of Eqn (2) for a uniform grid of source positions $\zeta_{ij}$.
For this calculation, each GPU thread independently computes the magnification for a given pixel in the map.
The map is then convolved on the GPU, by a number of radially-symmetric source intensity profiles of different
radii and fixed limb darkening ($\Gamma_I=0.53$). This produces a set of magnification maps corresponding to fixed $\rho$ values.  
We have adopted seven values of $\rho$ logarithmically-spaced over $(0.001,1)$. 
Our tests have shown that on currently available hardware it is faster to recompute these maps from scratch than to load them into
memory from hard-disk storage.
The magnification maps are loaded
into GPU texture memory for the next stage of the analysis -  a search over $(u_0, \phi, \rho, t_0, t_E)$ space for each given $(d,q)$.

\subsection{Re-parameterising $u_0$}

\begin{figure}
\includegraphics[width=\columnwidth]{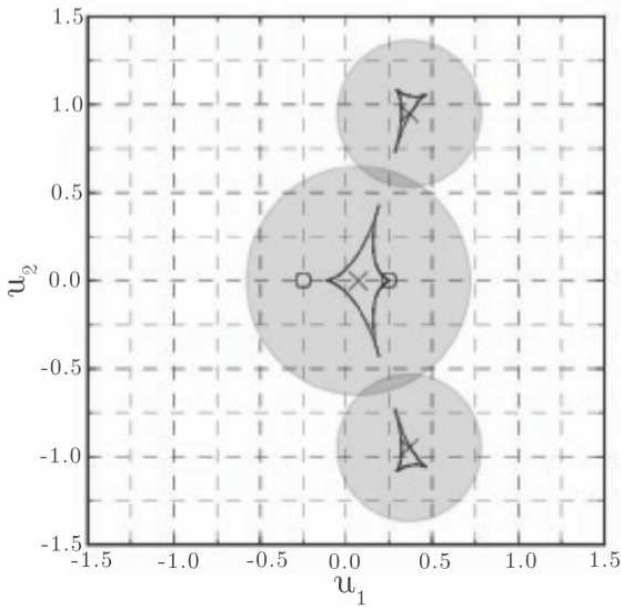}
\caption{Example of an $(r,\phi)$ search space about the caustics of a $d=0.7$, $q = 0.5$ binary. The coordinates $(u_1, u_2)$ are positions in the lens plane in Einstein radius units relative to the mid-point of the lens masses. Lens masses are marked by 'O' symbols and the centres of the search spaces are marked by 'X'.}
\label{fig:caustic_area_search_triple_a}
\end{figure}

Since $u_0$ is defined with respect to the centre of the binary lens, a search of trajectories in 
a magnification map over a grid of angles $\phi$ and impact parameters $u_0$ is liable to miss small caustic structures
that are located far from the centre. These small (planetary) caustics can be critical in identifying binary
lenses that include a low mass component.
With this in mind, we replace a search over $u_0$ with a search over
fixed radial distances, $r$, from the central regions of caustics. These radial distances are scaled by the sizes
of the caustics, as shown in Fig.~\ref{fig:caustic_area_search_triple_a}.

We conduct a search over $(t_0, t_E, \rho)$ for a fixed grid of $(r,\phi)$, where each $(r,\phi)$ pair is treated as a single
block on the GPU. We refer to an $(r,\phi)$ pair as a {\it trajectory}.
For this stage of the search, we use the simplex downhill method to locate a $\chi^2$ minimum.
At each iteration of the downhill search, each GPU thread is used to compute the contribution to $\chi^2$ from a 
single observed data point. A given $(t_i, t_0, t_E)$ maps to a point ($i$) on a linear trajectory defined by $(r,\phi)$, through
a given $(d,q,\rho)$ map. Bilinear interpolation of the magnification maps stored in texture memory is used to retrieve
the magnification $A(t_i)$ for the time of each data point.

\subsection{Optimisations}

There are several optimisations we make to improve the speed of the computation as follows.
\begin{enumerate}
\item
We use the observed peak-to-baseline magnitude range of the microlensing event to infer a minimum
magnification that needs to take place in order for a trial trajectory to be a viable solution. The magnification
as a function of position along a trajectory is quickly extracted from the point-source map and 
discarded if it doesn't fulfil this minimum magnification criterion.
\item
In cases where the lightcurve is obviously near in shape to that of a single lens (Paczy\'nski), we use the estimated
single lens parameters as starting points for $(t_0, t_E)$ in the simplex.
\item
In cases where the lightcurve has, or is expected to contain, more than one peak, we reject trajectories
where the point-source map contains fewer than the specified number of peaks. Since we are using a fixed grid of
$(d, q, r, \phi)$, we have pre-computed a library that specifies the number of peaks in each trajectory. A peak is considered to be any point on a trajectory where the magnification gradient changes from positive to negative.
\item
In non-Paczy\'nski cases where we are able to specify the time of a single peak, $t_0$ and $t_E$ are
related by 
\begin{equation}
t_0 = t_p - M_p t_E,
\label{EQN:1_peak_alignment}
\end{equation}
where $t_p$ is the input time of the peak from the data and $M_p$ is the distance along the trajectory of the peak 
from $t_0$ in units of Einstein radii.
$M_p$ is specified in our precomputed library. For a given trajectory, we can then adjust $t_0$ to best match the data in order to find our
starting simplex parameters.
\item
The optimum situation occurs where the times of multiple peaks can be provided. This allows $t_0$ and $t_E$ to be solved analytically 
for each viable trajectory since, in addition to Eqn.~\ref{EQN:1_peak_alignment}, 
\begin{equation}
t_E = \frac{t_{p2} - t_{p1}}{M_{p2}-M_{p1}} 
\label{EQN:2_peak_tE}
\end{equation}
where the $t_{p,i}$ are the set of specified times of peaks, and  $M_{p,i}$ are the pre-computed model peak times  in units of Einstein radii. 
Again, we use these values as starting points for the simplex search.
\end{enumerate}

\subsection{Results}

\begin{figure}
\includegraphics[width=\columnwidth]{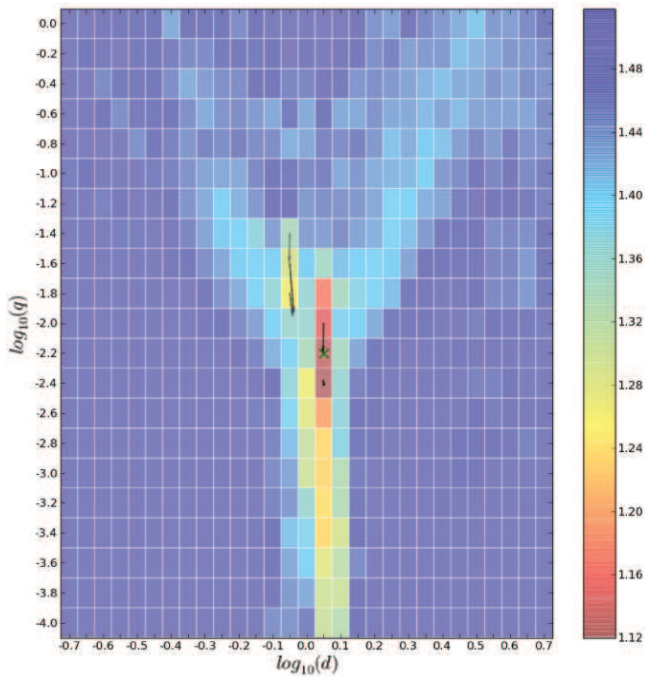}
\caption{The $\log \chi^2$ map in $(\log d,\log q)$ space for OGLE-2003-BLG-0235. Overlaid arrows show the 
	movement from seed solutions during the MCMC search.}
\label{fig:OB030235_HR_chi2_map}
\end{figure}

The outcome of our coarse grid search is a set of viable regions of parameter space where we expect to find the global solution.
Fig.~\ref{fig:OB030235_HR_chi2_map} shows an example of
coarse-search results projected onto $(\log d, \log q)$ space. 
Because of the coarseness of the $(\log d,\log q,r,\phi)$ grid, it is unreasonable to adopt the single best solution thus far found 
as being the likely seed for the global solution. Thus we generally choose a number of local minima from the coarse search for further
analysis. At this time, we adopt the three best wide solutions ($d \geq 1$) and narrow solutions ($d < 1$) as our seeds.  

\section{Refining solutions}
\label{Section:Refining_solutions}

To refine our seed solutions, we employ a Markov Chain Monte Carlo search. For this part of our modelling, we discard the 
magnification map approach and instead compute magnifications for a given set of trial parameters using an image-centred
ray shooting algorithm similar to that of \citet{Bennett2010}. The Markov chain sampling is implemented in the Python language
with CUDA C used to compute magnifications.

To implement our calculations on a GPU, we treat the epoch of each data point as a single block. Within a block, we use groups of threads
to perform integrations over images, with each thread corresponding to different locations on an image. 

Initially we compute the locations of the
caustics corresponding to the current $(d,q)$. This information is shared with all blocks (epochs). We then use 128 threads distributed over the
caustic edges to establish the proximity of the source to its nearest caustic. For source-caustic distances greater than $10\rho$ we
use the point-source approximation for the magnification, otherwise we proceed with image-centred ray shooting calculations as follows.

\begin{figure}
\includegraphics[width=\columnwidth]{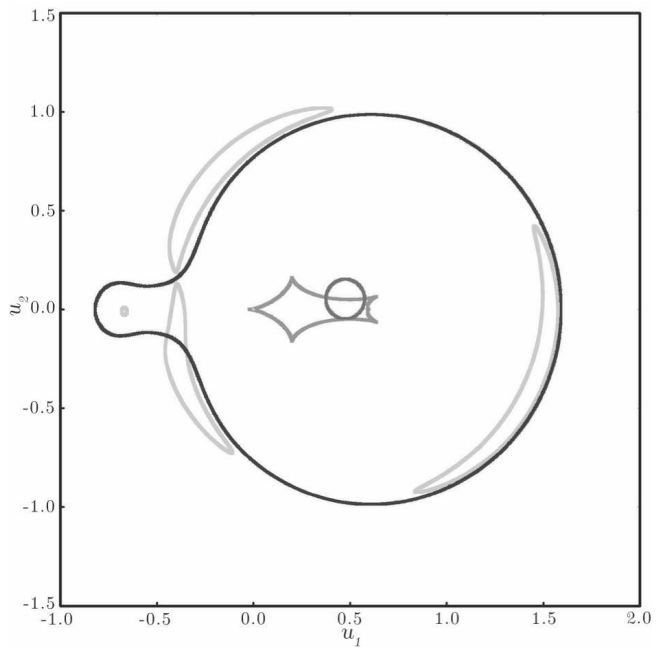}
\caption[Image around critical curves]{A finite source (dark central circle) passing over a caustic structure (five pointed central structure), resulting in 5 images (lightest grey) that closely follow the edge of the critical curves (darkest grey). This example also shows the merging of two images that cross the critical curve due to the source being partially inside of the caustic structure. }
\label{FIG:critical_curves}
\end{figure}

Again using 128 threads distributed around the source boundary and its centre, we compute image positions for these source points. 
Resampling near the highest magnification points is performed in order to detect situations where a source is grazing a caustic.
Starting from the centre of each image, a polar grid is iteratively expanded until it encompasses the entire image. Polar coordinates are used
since images are often crescent shaped, approximately following the locus of the critical curves (see Fig.~ \ref{FIG:critical_curves}). Tests are incorporated to detect 
merged images. 

We now integrate over each image with a weighting corresponding to the source intensity profile (found by inverse ray shooting).
To deal with the numerical problems introduced by the fact that the source intensity is zero at the boundaries, we follow the 
second-order integration scheme given by  \citet{Bennett2010}. 
Individual threads are used as grid points for the integration and we invoke parallel summation with loop unrolling
for rapid computation.

\subsection{Final solution with uncertainties}
\label{SubSection:Final_solution}

Upon completion of local area MCMCs, the best solution is chosen from the chain that produced the lowest final $\chi^2$ model, which we regard  as being close to the 
global minimum. From this solution the data errors are normalized by forcing $\chi^2$ per degree of freedom for each source of data to unity.
Although statistically undesirable, normalizing errors bars in this way prevents bias due to poorly-estimated uncertainties in particular data sources.
We then use the sampler EMCEE \citep{Foreman2013} along with CUDA image-centered ray shooting to determine our final solution and parameter covariances.

\subsection{Higher-order effects}
The inclusion of higher-order effects such as parallax and orbital motion can be incorporated in the GPU kernel to modify the source and lens
positions when performing inverse ray shooting, at minimal computational cost. However, new challenges arise 
when incorporating these effects into a search strategy whilst maintaining high-performance. Parallax (and xallarap) could, in principle, be 
added to the current grid search strategy, as they only affect the interpolated source trajectory. The inclusion 
of orbital motion at this stage is not possible since multiple $(d, q)$ magnification maps per trajectory would be required. These 
higher-order effects are typically only considered if necessary after the original seven parameters have been determined, at 
which point they are incorporated into the EMCEE search, which uses image-centred ray shooting.

\section{Examples}
\label{Section:Examples}

As demonstrations and verification of our code, we present results below for two well know events that have been previously analysed.

\subsection{OGLE-2003-BLG-0235}

This was the first microlensing event in which a planet was discovered, see \citet{Bond2004}. It has more recently been reanalysed by \citet{Bennett2010}.
Data was taken by the OGLE and MOA survey telescopes that show a moderate magnification microlensing event ($A_0 \sim 5.5$)
that peaked at around JD 2452847. The event shows a caustic passage that lasted for $\sim 7$ days from JD 2452834.
In Table~\ref{TAB:modelsogl-2003-blg-0235} we show the solutions found by 
\citet{Bond2004}.

\begin{table*}
\centering
\caption{Binary lens seven parameter model maximum likelihood solutions of MOA-2003-BLG-0053/OGLE-2003-BLG-0235, showing planetary solutions and the best non-planetary model \citep{Bond2004} and from our code, based on EMCEE runs using error-normalized data.}
\label{TAB:modelsogl-2003-blg-0235}
\begin{tabular}{|c|c|c|c|c|c|}
\hline
& \multicolumn{3}{|c|}{\citet{Bond2004} solutions} & \multicolumn{2}{|c|}{This paper} \\
			& Best				& Early caustic	& Best				& Wide		& Close \\ 
Parameter	& planetary			& planetary	& non-planet	& planetary	& planetary\\ 
\hline
$d$		& $1.120 \pm 0.007$			& $1.121$			& $1.090$		 &$1.117 $			& $0.9143$\\
$q (\times 10^-3)$		& $3.9 \substack{+1.1 \\ -0.7} $		& $7.0$		& $30.0$	 & $6.5$			& $11.5$\\
$\rho (\times 10^-4)$	& $9.6 \pm 1.1$		& $10.4$		& $8.8$	 & $9.9$ 			& $8.6$\\
$u_0$	& $0.133 \pm 0.003$			& $0.140$			& $0.144$		 & $0.138$			& $0.154$\\
$\phi$	& $0.7644 \pm 0.0024$		& $0.6789$		& $0.1379$	 & $0.677$			& $3.25$\\
$t_0$	& $2848.06 \pm 0.13$		& $2847.90$		& $2846.20$	 & $2847.91$			& $2850.02$\\
$t_E$	& $61.5 \pm 0.18$			& $58.5$			& $57.5$		 & $59.5$	  			& $56.8$\\ \hline
$\chi^2/dof$	& 			& 		& 	 & 1508.19/1374			&  1597.67/1374 \\ 
\hline
\end{tabular}
\end{table*}

\begin{figure}
\includegraphics[width=\columnwidth]{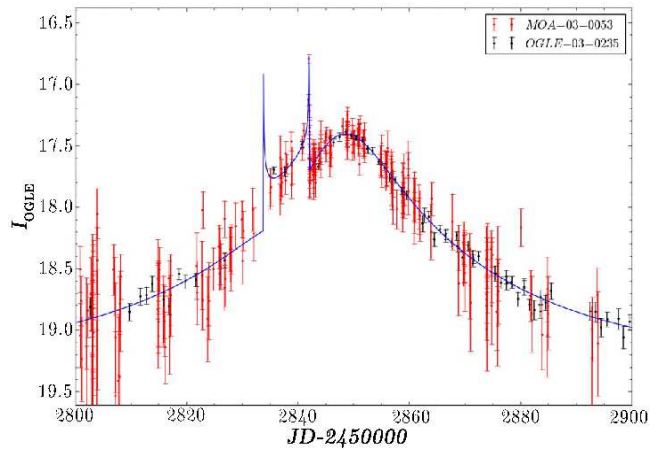}
\caption{Final fitted lightcurve for the marginalized wide solution of OGLE-2003-BLG-235}
\label{fig:ob120235_final}
\end{figure}

We have modelled the event using the OGLE and MOA data from \citet{Bond2004} that has been made available through the
NASA Exoplanet Archive\footnote{http://exoplanetarchive.ipac.caltech.edu/}. 
The entire OGLE data set of 285 points was used, while the MOA data set was reduced from 1152 data points down to 
1089 by rejecting points with very large uncertainties.
We specified a minimum base-peak lightcurve amplitude of 2.0 mag, and demanded that solutions contained
at least three lightcurve peaks, with two of these located near JD2452842.08 (the caustic exit) and JD2452848.0 (the central peak).
The $\chi^2$ map from our initial grid search is shown in Fig~\ref{fig:OB030235_HR_chi2_map}. This initial search took 59 minutes using 
an NVIDIA Tesla C2075 GPU card. MCMC models using the image-centred ray shooting procedures discussed in Section~\ref{Section:Refining_solutions}
were then started from the three best close ($d < 1$) and wide 
($d >= 1$) solutions. Well-converged solutions, using a total of 4000 iteration steps per chain  were obtained for these 6 chains in around 1.85 hours.
Arrows on Fig.~\ref{fig:OB030235_HR_chi2_map} indicate the movement of solutions during the 
MCMC iterations. 
In Table~\ref{TAB:modelsogl-2003-blg-0235}  we show the parameters of our two best solutions
compared with those found by \citet{Bond2004}. 
Our final solution, incorporating normalized data uncertainties, and with parameter uncertainties generated using EMCEE, 
is given in Table~\ref{TAB:my_ogl-2003-blg-0235} and Fig. \ref{fig:ob120235_final}, and is the same as the best solutions found previously. 

This analysis also serves as a performance comparison with the method of \citet{Bennett2010}, who achieved the same `Best planetary' solution of Table~\ref{TAB:modelsogl-2003-blg-0235} by searching a restricted parameter space of slightly more that $70,000$ parameter sets in 5 hours and 14 minutes. 
Using our methods, we determined our 6 best coarse solutions by exploring an unrestricted parameter space of $8,197,199$ parameter sets in 59 minutes.
This includes the time for each parameter set to potentially perform a simplex downhill search of up to $5000$ steps over $(t_0, t_E, \rho)$
including the subsequent 6 MCMC explorations, the analysis explored between 8 million and 40 million parameter trials in 2 hours and 50 minutes.

\begin{table}
\caption{Final seven parameter binary lens model solutions of OGLE-2003-BLG-0235 from error-normalized data. These are the marginalized posterior mean and standard deviation for each parameter.}
\label{TAB:my_ogl-2003-blg-0235}
\begin{tabular}{|c|l|l|}
\hline
		 & Marginalized	&	Marginalized	\\
Parameter & wide solution	&	close solution	\\ \hline
$d$		& $	1.120 \pm	0.004$		&	$ 0.915 \pm 0.002$		\\
$q (\times 10^{-3})$	& $ 6.3 \pm 0.6$	&	$ 11.7\pm 1.0$		\\
$\rho (\times 10^{-4})$	& $ 10.0 \pm 1.2$	& $ 8.98 \pm 0.71$		\\
$u_0$	& $	0.137 \pm 0.004$		&	$ 0.154 \pm 0.005$		\\
$\phi$	& $	0.695	\pm	0.024$	&	$ 3.24 \pm 0.02	$	\\
$t_0$	& $	2847.91	\pm	0.11$	&	$ 2850.02 \pm 0.13$		\\
$t_E$	& $	59.8	\pm	1.6$			&	$ 56.8 \pm 1.5$		\\ \hline
\end{tabular}
\end{table}

\subsection{OGLE-2012-BLG-0406}

This is an example of an event detected and monitored by the OGLE IV survey group, with intensive follow-up
observations by several other groups. Following detection in 2012 April, the event was predicted to have a low peak magnification and was 
considered low priority for follow-up. An anomalous brightening was recognised in early July, prompting monitoring by the PLANET and 
microFUN groups using telescopes at SAAO and CTIO. Following a peak in the lightcurve on July 5 and recognition by C. Han that the lightcurve 
was probably planetary in nature, several other telescopes began to intensively monitor the event.

The lightcurve has been modelled by \citet{Poleski2014} using only OGLE survey telescope data, and 
\citet{Tsapras2014} who used data from ten separate observing sites.
Both of these papers used a similar initial search routine, with a few differences in the filtering and error scaling of the data used.  \citet{Poleski2014} assumed a limb-darkening coefficient of $\Gamma_I=0.353$ whereas  \citet{Tsapras2014} assumed $\Gamma_I=0.53$. 

Due to the deviations from a point-source lightcurve being of short time scale on the shoulder of a Paczy\'nski curve, it is safe to conclude that the cause of the anomaly is from the source passing close to a central caustic. In such a case, the single lens parameters of  $u_0$, $t_0$, and $t_E$ (which can be determined with relative ease) are similar to the matching parameters of the binary lens model. \citet{Tsapras2014} used this information to perform a hybrid grid search of $d$, $q$, and $\phi$, where $u_0$, $t_0$, $t_E$, and $\rho$ are minimized using MCMC methods. The grid was searched between the limits of $-1 \leq \log_{10}(d) \leq1$, $-5 \leq \log_{10}(q) \leq1$, and $0 \leq \phi \leq 2\pi$ with the aim to locate all areas of local minimum, before further refinement by narrowing the grid search parameter space. Finally a $\chi^2$ optimization in all seven parameters was performed in each local minimum to determine the best solution. The \citet{Tsapras2014} and  \citet{Poleski2014} binary lens model solutions are shown in Table \ref{TAB:ob120406}.

\begin{figure}
\includegraphics[width=\columnwidth]{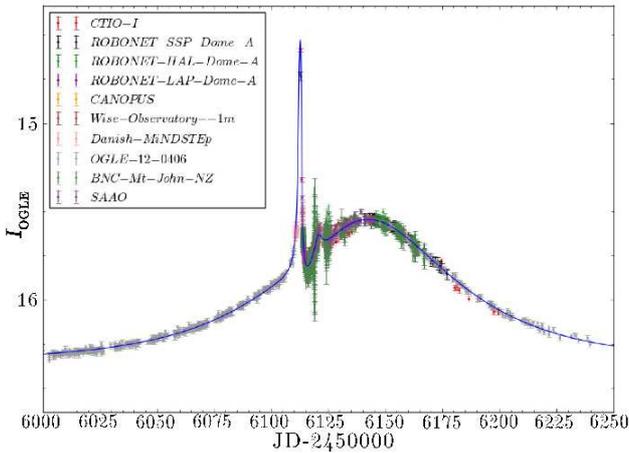}
\caption{Final fitted lightcurve for OGLE-12-BLG-406}
\label{fig:My_all_light curve_ob120406}
\end{figure}

We have used our code as described in Sections 3-5 above to fit the data sets used by \citet{Poleski2014}  (OGLE only) and by \citet{Tsapras2014}. 
The \citet{Tsapras2014} data consists of 81 observations from CTIO I band Chile (microFUN), 181 from Robonet Siding Spring Australia, 83 from Robonet Haleakala Hawaii, 131 from Robonet La Palma,  210 from Canopus Observatory Tasmania Australia (PLANET), 180 from Wise Observatory Israel, 473 from Danish 1.5-m La Silla Chile (MiNDSTEp), 3013 from OGLE Las Campanos Chile, 1856 from the B\&C 60-cm Mt John New Zealand (MOA), and 226 from SAAO 1-m South Africa (PLANET). Our initial grid searches located the same areas of local minima as \citet{Poleski2014}  and \citet{Tsapras2014}. 
Our subsequent MCMC refinements using image-centred ray shooting determined the same minima as those previous solution as 
given in Table~\ref{TAB:ob120406}. The fitted multi-data-set light curve is shown in  Fig.~\ref{fig:My_all_light curve_ob120406}.

\begin{table*}
\centering
\caption{Seven-parameter binary lens model solutions for event OGLE-2012-BLG-0406.}
\label{TAB:ob120406}
\begin{tabular}{|c|l|l|l|l|}
\hline
& \multicolumn{2}{|c|}{All data} & \multicolumn{2}{|c|}{OGLE only} \\
Parameter & 	\citet{Tsapras2014}  	&    This paper &  \citet{Poleski2014}  & This paper           \\ 
\hline
$d$       	& $1.346 \pm 0.001$     			& $1.3466 \pm 0.0009$        	 		& $1.3500 \pm 0.0016$     		&   $1.3505 \pm 0.0019$  \\
$q (\times 10^{-3} )$       	& $5.33 \pm 0.04$ 	& $5.32 \pm 0.04$ 	& $5.78 \pm 0.08$	&  $5.78 \pm 0.09$\\
$\rho (\times 10^{-2})$    	& $1.103 \pm 0.008 $& $1.086 \pm 0.011$ 	& $0.98 \pm 0.05$ 	&   $0.95 \pm 0.05$\\
$u_0$     	& $0.532 \pm 0.001$      			& $0.5314 \pm 0.0013$ 				& $0.5425 \pm 0.0022$    			&   $0.540 \pm 0.003$   \\
$\phi$    	& $0.852 \pm 0.001$         			& $0.8555 \pm 0.0013$      			& $0.8653 \pm 0.0016$   			& $0.8654 \pm 0.0017$  \\
$t_0$     	& $6141.63 \pm 0.04$       		& $6141.46 \pm 0.04$ 				& $6141.59 \pm 0.03$ 			&   $6141.43 \pm 0.05$    \\
$t_E$     	& $62.37 \pm 0.06$       			& $62.52 \pm 0.12$     				& $62.63 \pm 0.16$     			&     $62.52 \pm 0.19$   \\ 
\hline
\end{tabular}
\end{table*}

\section{Summary}
\label{Section:Summary}

In this paper we have described a fast method for fitting binary gravitational microlensing events. The code is
written in CUDA C and Python, and the computationally intensive parts are run on massively parallel GPU
hardware. The code will scale naturally to higher performance as GPU technology evolves.

As part of our algorithms, we introduce several optimisations that improve the ability to detect features due
to small caustics and the speed of execution. Results from our ongoing analysis of current events are
reported through our website \footnote{http://www2.phys.canterbury.ac.nz/$\sim$u-lenser/}.

\section{Acknowledgements}
MDA thanks the Department of Astronomy, University of Wisconsin-Madison, for hospitality during the writing of
this paper. We thank NVIDIA Corporation for supporting our work.

\bsp	
\label{lastpage}
\end{document}